\documentclass[%
preprint,
amsmath,amssymb,
aps,
pra,
]{revtex4-1}
\usepackage{graphicx}
\usepackage{bm}
\usepackage{multirow}
\usepackage{rotating}
\usepackage{float}
\usepackage{dcolumn}
\usepackage{amsmath}
\usepackage{amssymb}
\usepackage{enumerate}
\usepackage[shortlabels]{enumitem}

\begin{document}

\title{Theoretical assessment of boron incorporation in nickel ferrite under conditions of operating nuclear pressurized water reactors (PWRs)}

\author{Zs. R\'{a}k}
\email{zrak@ncsu.edu}
\author{E. W. Bucholz}
\author{D. W. Brenner}
\affiliation{Department of Materials Science and Engineering, North Carolina State University, Raleigh, 27695-7907}

\date{\today}

\begin{abstract}

A serious concern in the safety and economy of a pressurized water nuclear reactor is related to the accumulation of boron inside the metal oxide (mostly NiFe${}_{2}$O${}_{4}$ spinel) deposits on the upper regions of the fuel rods. Boron, being a potent neutron absorber, can alter the neutron flux causing anomalous shifts and fluctuations in the power output of the reactor core. This phenomenon reduces the operational flexibility of the plant and may force the down-rating of the reactor. In this work an innovative approach is used to combine first-principles calculations with thermodynamic data to evaluate the possibility of B incorporation into the crystal structure of NiFe${}_{2}$O${}_{4}$, under conditions typical to operating nuclear pressurized water nuclear reactors. Analyses of temperature and pH dependence of the defect formation energies indicate that B can accumulate in NiFe${}_{2}$O${}_{4}$ as an interstitial impurity and may therefore be a major contributor to the anomalous axial power shift observed in nuclear reactors. This computational approach is quite general and applicable to a large variety of solids in equilibrium with aqueous solutions.

\end{abstract}

\pacs{}

\maketitle 

\section{\label{intro}Introduction}

A major impediment that prevents pressurized water reactors (PWRs) from operating at higher duty and longer cycles is the accumulation of boron within metal oxide scales that deposit on the upper spans of fuel assemblies. Boron, in the form of boric acid (H${}_{3}$BO${}_{3}$), is added to PWR coolant to control the neutron flux while lithium hydroxide (LiOH) is also dosed to control the acidity of the coolant and to reduce corrosion. Because of the large neutron absorption cross section of ${}^{10}$B, a small amount of accumulated B is sufficient to cause an abnormal decrease in the neutron flux, which shifts the power output toward the bottom half of the reactor core. This phenomenon, known as axial offset anomaly (AOA), has been observed in high-duty reactors that run long fuel-cycles. In extreme form, AOA can decrease the reactor shutdown margin sufficiently to force major power reduction leading to substantial economic losses~\cite{epri_1, epri_2}.

AOA modeling efforts traditionally assume that boron deposition within the metal oxide scales (which are commonly referred to as CRUD, an acronym for Chalk River Unidentified Deposits) occurs predominantly through precipitation of lithium borate compounds such as LiBO${}_{2}$, Li${}_{2}$BO${}_{7}$, Li${}_{2}$B${}_{4}$O${}_{7}$~\cite{epri_2, epri_3, epri_4, epri_5, epri_6, uchida2011}. Although the retrograde solubility of these borates could explain the ``lithium return'' experienced during plant shutdown, they have never been observed experimentally in PWR CRUD. Using M\"{o}ssbauer spectroscopy together with XRD on CRUD scrapes recovered from high duty PWRs, Sawicki identified the precipitation of Ni${}_{2}$FeBO${}_{5}$ as a possible mechanism for B deposition~\cite{sawicki2008,sawicki2010}. Mesoscale CRUD models developed by Short \textit{et al.}, assume that supersaturation of boric acid leads to precipitation of boron trioxide (B${}_{2}$O${}_{3}$) within the CRUD~\cite{short2013}.

In recent work, we combined \textit{ab initio }calculations with experimental formation enthalpies to investigate the incorporation of B into the structure of nickel ferrite (NiFe${}_{2}$O${}_{4}$, NFO) as a potential new mechanism for B deposition within CRUD~\cite{rak2014}. Assuming solid-solid (and solid-gas) equilibrium between nickel ferrite and elemental reservoirs of Fe, Ni, B (and O${}_{2}$ gas) we found that it is thermodynamically favorable for B to form secondary phases with Fe, Ni, and O (\textit{e. g.} B${}_{2}$O${}_{3}$, Fe${}_{3}$BO${}_{5}$, and Ni${}_{3}$B${}_{2}$O${}_{6}$) instead of entering the NFO structure as a point defect. Building on previous works~\cite{o'brien2013, o'brien2014, matsunaga2008}, the present study attempts to deal with the same question, however, here the defect formation energies are evaluated assuming solid-liquid equilibrium between NFO and the surrounding aqueous solution of Ni, Fe and dissolved boric acid (H${}_{3}$BO${}_{3}$). To set up solid-liquid equilibrium, the chemical potentials of individual aqueous species are defined as a function of temperature, pressure, and concentration and are linked to the chemical potentials of the ionic species in solid. This new scheme allows for the evaluation of defect formation energies under conditions that are specific to operating nuclear PWRs. The approach is quite general and applicable to a large variety of solids in equilibrium with aqueous solutions.

\section{\label{defect_formation_ene}Defect formation energies under solid-liquid equilibrium}

The first-principles calculations required for the present study have been carried out within the Density Functional Theory (DFT) using the same computational parameters and crystal models that are specified in Ref.~\onlinecite{rak2014}.

The formation of a defect in a crystalline solid can be regarded as an exchange of particles (atoms and electrons) between the host material and chemical reservoirs. The formation energy of a defect \textit{D} in charge state \textit{q} can be written as~\cite{zhang1991,zhang2002}

\begin{equation} \label{eq1}
\Delta H_{f} \left(D^{q} \right)=E\left(D^{q} \right)-E_{0} +\sum _{i}n_{i} \mu _{i} +q \left(E_{F} +E_{VBM}^{def} \right).
\end{equation}

In Eq. \eqref{eq1}, $E\left(D^{q} \right)$ and $E_{0} $ are the total energies of the defect-containing and defect free solids, calculated within the DFT. The third term on the right side of  Eq. \eqref{eq1} represents the change in energy due to the exchange of atoms between the host compound and the chemical reservoir, where $\mu _{i}$ is the atomic chemical potential of the constituent \textit{i} (\textit{i} = Ni, Fe, or B). The quantity $n_{i}$ represents the number of atoms added to $\left(n_{i} <0\right)$ or removed from $\left(n_{i} >0\right)$ the supercell. The quantity $E_{F}$ is the Fermi energy referenced to the energy of the valence band maximum (VBM) of the defective supercell, $E_{VBM}^{def} $. This value is calculated as the VBM energy of the pure NFO, corrected by aligning the core potential of atoms far away from the defect in the defect-containing supercell with that in the defect free supercell~\cite{zhang2002}. The quantity $q$ represents the charge state of the defect, \textit{i. e.} the number of electrons exchanged with the electron reservoir with chemical potential $E_{F}$.

Under solid-liquid equilibrium conditions, the chemical potentials of the ionic species in the solid,$\mu _{i^{z_{i} } } $, are equal to the chemical potential of the aqueous species in the saturated solution, $\mu _{i^{z_{i} } ,aq} $. To derive an expression for defect formation energy, Eq.~\eqref{eq1} has to be written in terms of ionic chemical potentials, $\mu _{i^{z_{i} } } $, instead of atomic chemical potentials, $\mu _{i}$. This can be accomplished by adding and subtracting the term $\sum _{i}n_{i} z_{i} \left(E_{F} +E_{VBM}^{0} \right) $ from Eq. \eqref{eq1}. This term can be interpreted as the energy necessary to exchange electrons between the electron reservoir and the atomic species in the pure NFO. If we combine this energy with the atomic chemical potential (third term in Eq.\eqref{eq1}) we obtain the ionic chemical potential of species in NFO:

\begin{equation} \label{eq2}
\sum _{i}n_{i}  \mu _{i} +\sum _{i}n_{i} z_{i} \left(E_{F} +E_{VBM}^{0} \right) =\sum _{i}n_{i}  \mu _{i^{z_{i} } } .
\end{equation}

In Eq. \eqref{eq2} $z_{i} $ represents the ionic charge, \textit{i. e}. the number of electrons exchanged with the electron reservoir to create the ionic species in NFO, and $E_{VBM}^{0}$ is the energy of the VBM in the pure NFO to which the energy of the electron reservoir $\left(E_{F} \right)$ is referenced. Using Eq.~\eqref{eq2} together with the solid-liquid equilibrium condition, $\left(\mu _{i^{z_{i} } } =\mu _{i^{z_{i} } ,aq} \right)$, the defect formation energy becomes:

\begin{equation} \label{eq3}
\Delta H_{f} \left(D^{q} \right)=\Delta E\left(D^{q} \right)+\sum _{i}n_{i} \mu _{i^{z_{i} } ,aq}  -\sum _{i}n_{i} z_{i} \left(E_{F} +E_{VBM}^{0} \right) +q\left(E_{F} +E_{VBM}^{def} \right)
\end{equation}
where $\Delta E\left(D^{q} \right)$ is the energy difference between the defect containing and defect free supercells. Therefore, to calculate the defect formation energy, the chemical potentials of the aqueous species have to be evaluated. The scheme described above has the advantage that it decouples the ionic charge from the charge state of the defect; charge neutrality is achieved through exchange of electrons with the electron reservoir with energy equal to the Fermi level.

\section{\label{aqueous_chem_pot}Aqueous chemical potentials}

The chemical potential of the aqueous ions, $\mu _{i^{z_{i} } ,aq}$, can be written as the sum of the standard chemical potential, $\mu _{i^{z_{i} } ,aq}^{0} $,  and a temperature dependent term

\begin{equation} \label{eq4}
\mu _{i^{z_{i} } ,aq} =\mu _{i^{z_{i} } ,aq}^{0} +RT\ln a_{i^{z_{i} } }  .
\end{equation}

In Eq.~\eqref{eq4}, $R{\rm \; =\; 8.314\; J/K}\cdot {\rm mol}$ is the universal gas constant and $a_{i^{z_{i} } }$ is the activity of the ionic species $i^{z_{i} }$ in the aqueous solution. In the present case, because NFO is weakly soluble in water, the activity of the ionic species can be approximated by the concentration of ions in the solution.

The standard chemical potentials of aqueous cations can be evaluated using thermochemical data combined with theoretical total energies calculated within the DFT framework. If we consider the reaction

\begin{equation} \label{eq5)}
M(solid)+zH^{+} (aq)\to M^{z+} (aq)+\frac{z}{2} H_{2} (gas),
\end{equation}
the equilibrium condition can be written as

\begin{equation} \label{eq6)}
\mu _{M,solid}^{0} +z\mu _{H^{+} ,aq}^{0} =\mu _{M^{z+} ,aq}^{0} +\frac{z}{2} \mu _{H_{2} ,gas}^{0}
\end{equation}
Therefore, $\mu _{M^{z+} ,aq}^{0}$ can be expressed using the standard Gibbs energy of formation of ions in aqueous solution as

\begin{equation} \label{eq7}
\mu _{M^{z+} ,aq}^{0} =\Delta G_{f}^{0} \left(M^{z+} ,aq\right)+\mu _{M,solid}^{0} +z\left(\mu _{H^{+} ,aq}^{0} -\frac{1}{2} \mu _{H_{2} ,gas}^{0} \right).
\end{equation}

The Gibbs energies of formation required for Eq. \eqref{eq7} are obtained from that SUPCRT database~\cite{windman2008, johnson1992} which uses the revised Helgeson-Kirkham-Flowers equation of state to predict the thermodynamic behavior of aqueous species at high temperature and pressure~\cite{tanger1988}.

The chemical potentials of the solid phases $\mu _{M,solid}^{0}$ are usually approximated by the total energy per atom of the elemental solid calculated within the DFT framework. However, as pointed out in earlier work, this approach suffers from incomplete error cancellation when total DFT energies of physically and chemically dissimilar systems are compared~\cite{jones1989, stevanovic2012, lany2008, jain2011}. Therefore, to compute the elemental-phase chemical potentials of the Fe, Ni, B, and O, we extend the database of 50 elemental energies published by Stevanovic \textit{et al.}~\cite{stevanovic2012} to include B. To do this we add 26 B-containing binaries to the large fitting set of 252 compounds that have been used by Stevanovic \textit{et al.} and solve the overdetermined system of 278 equations for 51 elements using a least-square approach, as described in Refs.~\onlinecite{stevanovic2012,lany2008}. The calculated DFT energies and experimental formation enthalpies of the 26 B-containing binaries are listed in Table~\ref{table1}, while the 51 elemental-phase chemical potentials are given in Table~\ref{table2} in the Appendix.

\begin{table}[!htbp]
\caption{\label{table1} DFT energies and experimental enthalpies of formation of 26 B-containing binaries that have been added to the fitting set in Ref.~\onlinecite{stevanovic2012}, to calculate the elemental-phase chemical potentials. Theoretical enthalpies of formation are also listed.}
\begin{ruledtabular}
\begin{tabular}{l c c c }
Compound & DFT energy (eV/atom) & $\Delta H^{exp } $ (eV/atom) & $\Delta H^{theor} $ (eV/atom) \\ \hline
B${}_{2}$S${}_{3}$ & -5.58 & -0.52 & -0.56 \\
B${}_{2}$O${}_{3}$ & -8.02 & -2.64 & -2.57 \\
B${}_{6}$O & -7.14 & -0.78 & -0.84 \\
BN & -8.79 & -1.31 & -1.32 \\
BP & -6.45 & -0.60 & -0.44 \\
Cr${}_{3}$B${}_{4}$ & -7.10 & -0.42 & -0.30 \\
CrB & -7.12 & -0.41 & -0.27 \\
CrB${}_{2}$ & -8.00 & -0.43 & -1.25 \\
Fe${}_{2}$B & -6.37 & -0.35 & -0.12 \\
FeB & -6.50 & -0.38 & -0.17 \\
HfB${}_{2}$ & -8.07 & -1.14 & -1.18 \\
MgB${}_{2}$ & -5.11 & -0.32 & -0.36 \\
MgB${}_{4}$ & -5.80 & -0.22 & -0.33 \\
Mn${}_{2}$B & -6.85 & -0.32 & -0.10 \\
Mn${}_{3}$B${}_{4}$ & -6.96 & -0.35 & -0.28 \\
MnB & -7.01 & -0.37 & -0.30 \\
MnB${}_{2}$ & -6.95 & -0.33 & -0.29 \\
NbB${}_{2}$ & -7.45 & -0.61 & -0.78 \\
Ni${}_{2}$B & -4.96 & -0.22 & -0.32 \\
Ni${}_{3}$B & -4.67 & -0.23 & -0.27 \\
Ni${}_{4}$B${}_{3}$ & -5.25 & -0.27 & -0.34 \\
NiB & -5.44 & -0.24 & -0.32 \\
TaB${}_{2}$ & -8.12 & -0.72 & -0.76 \\
TiB & -6.91 & -0.83 & -0.84 \\
TiB${}_{2}$ & -7.30 & -1.09 & -1.06 \\
ZrB${}_{2}$ & -7.58 & -1.12 & -1.20 \\
\end{tabular}
\end{ruledtabular}
\end{table}

The last term in parenthesis in Eq.~\eqref{eq7} can be evaluated using a Born-Haber-type cycle of hydrogen. The formation of an aqueous proton in water, $H^{+} \left(aq\right)$, is described by the reaction $ {1}/{2} H_{2} \left(gas\right)\to H^{+} \left(aq\right)+e^{-} \left(gas\right)$. The path of this reaction can be envisioned as dissociation of the H${}_{2}$ molecule followed by ionization of the H atom and dissolution of the proton in water. Therefore, the quantity required in Eq. \eqref{eq7} can be calculated as the sum of free energies of atomization, ionization, and solvation of H. All of these free energies are available in the literature and the free energy of formation of aqueous proton has recently been evaluated at 298K and 1 bar as 4.83 eV~\cite{todorova2014}. This value is used in the present study.

To evaluate the chemical potential of aqueous species (Eq.\eqref{eq4}) and the defect formation energies (Eq. \eqref{eq3}), the remaining quantity that has to be determined is the ionic concentration in aqueous solution.  This could be obtained assuming a saturated solution with respect to NFO and using experimental solubility data combined with charge neutrality requirements. However, because the experimental data related to NFO solubility is limited, in the present work we use [Fe${}^{3+}$] = 4.17$\times$10${}^{-13}$ and [Ni${}^{2+}$] = 1.66$\times$10${}^{-14}$ mol/dm${}^{3}$, concentrations that are characteristic to an operating PWR~\cite{epri_3, epri_4}.

To assess the validity of the scheme described above regarding the standard chemical potentials as well as the predictive power of the elemental-phase chemical potentials listed in Table~\ref{table2}, we evaluate the Gibbs free energies of formation of NiO and NiFe${}_{2}$O${}_{4 }$based on the following reactions:

\begin{equation} \label{eq8}
Ni^{2+} +H_{2} O\to NiO+2H^{+}
\end{equation}

\begin{equation} \label{eq9}
Ni^{2+} +2Fe^{3+} +4H_{2} O\to NiFe_{2} O_{4} +8H^{+}
\end{equation}

Under equilibrium conditions, employing Eq. \eqref{eq7} to express the chemical potentials of Ni${}^{2+}$ and Fe${}^{3+}$, and considering that $\mu _{H_{2} }^{0} +1/2 \mu _{O_{2} }^{0} =\mu _{H_{2} O}^{0}$ (with $1/2 \mu _{O_{2} }^{0} =\mu _{O}^{0}$), the change in Gibbs energies of reactions \eqref{eq8} and \eqref{eq9} are:

\begin{equation} \label{eq10}
\Delta G^{0} \left(NiO\right)=\mu _{NiO,s}^{0} -\Delta G_{f}^{0} \left(Ni^{2+} ,aq\right)-\Delta G_{f}^{0} \left(H_{2} O\right)-\mu _{Ni,s}^{0} -\mu _{O}^{0}
\end{equation}

\begin{equation} \label{eq11}
\begin{array}{l} {\Delta G^{0} \left(NiFe_{2} O_{4} \right)=} \\ {=\mu _{NiFe_{2} O_{4} ,s}^{0} -\Delta G_{f}^{0} \left(Ni^{2+} ,aq\right)-2\Delta G_{f}^{0} \left(Fe^{3+} ,aq\right)-4\Delta G_{f}^{0} \left(H_{2} O\right)-\mu _{Ni,s}^{0} -2\mu _{Fe,s}^{0} -4\mu _{O}^{0} } \end{array}
\end{equation}

\begin{figure}[t]
\centering
\includegraphics[scale=0.35]{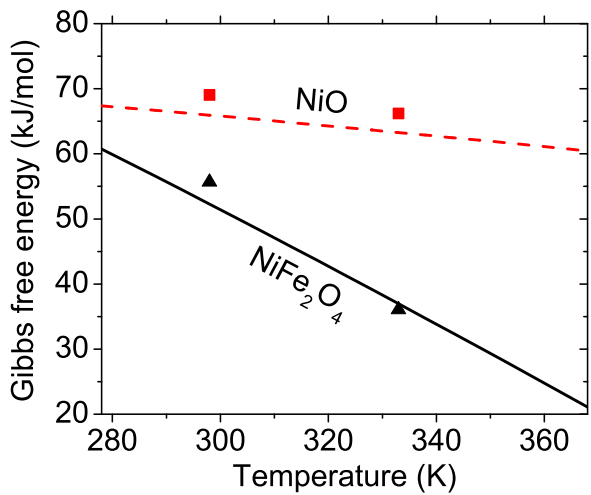}
\caption{(Color online) Gibbs free energies of formation of NiO and NFO from aqueous ions. The experimental values, indicated by squares (for NiO) and triangles (for NFO) at 298 and 333 K, are in good agreement with the calculated values.}
\label{fig1}
\end{figure}

In Eqs. \eqref{eq10} and \eqref{eq11} $\mu _{NiO,s}^{0}$ and $\mu _{NiFe_{2} O_{4} ,s}^{0}$ represent the chemical potentials of solid NiO  and NiFe${}_{2}$O${}_{4}$ and they are approximated by the total DFT energies per formula unit (f. u.). The Gibbs free energies of formation of aqueous Ni${}^{2+}$ and Fe${}^{3+}$ as well as the free energy of water are taken from the SUPCRT database~\cite{windman2008, johnson1992}, while the elemental-phase chemical potentials of solid Ni, Fe and gaseous O are listed in Table~\ref{table2}. The reaction energies at 0.1 MPa, calculated using Eqs. \eqref{eq10} and \eqref{eq11}, are plotted as a function of temperature in Fig.~\ref{fig1}. The calculated values are in good agreement with the available experimental free energies at 298 and 333 K~\cite{fujiwara2004}, as illustrated in Fig.~\ref{fig1}. In general the experimental values $\left(\Delta G_{298}^{0} \left(NiO\right)=69.04\right. $, $\Delta G_{333}^{0} \left(NiO\right)=66.19$, and $\left. \Delta G_{298}^{0} \left(NiFe_{2} O_{4} \right)=55.66{\rm \; kJ/mol}\right)$ are slightly higher than the calculated ones $\left(\Delta G_{298}^{0} \left(NiO\right)=65.95\right. $, $\Delta G_{333}^{0} \left(NiO\right)=63.28$, and $\left. \Delta G_{298}^{0} \left(NiFe_{2} O_{4} \right)=52.29{\rm \; kJ/mol}\right)$, with the exception of NFO at 333K where the agreement between experimental $\left(\Delta G_{333}^{0} \left(NiFe_{2} O_{4} \right)=36.04{\rm \; kJ/mol}\right)$ and theoretical values $\left(\Delta G_{333}^{0} \left(NiFe_{2} O_{4} \right)=36.96{\rm \; kJ/mol}\right)$ is remarkably good. This indicates that the above method is adequate for evaluation of aqueous chemical potentials and free energies of reaction and formation.

To evaluate the possibility of B incorporation into the crystalline structure of CRUD, we compute the formation energies of B-related point defects in NFO, assuming solid-liquid equilibrium based on the method described above. The point defects investigated here are substitutional and interstitial B impurities. The crystal and defect models used in the present work are identical to those described in Ref.~\onlinecite{rak2014}. Three types of substitutional defects are possible: B can substitute for a tetrahedral or octahedral Fe atom $\left(B_{Fe^{T} } {\rm \; or\; }B_{Fe^{O} } \right)$ or it can occupy an octahedral Ni site $\left(B_{Ni^{O} } \right)$. Similarly, there are three different interstitial sites in the spinel structure that can be occupied by B: one octahedral site $\left(B_{O} \right)$ and two tetrahedral sites$\left(B_{T_{1} } {\rm \; and\; }B_{T_{2} } \right)$. While both T${}_{1}$ and T${}_{2}$ are tetrahedrally coordinated by O atoms, the T${}_{1}$ site has four nearest neighbor (NN) cations and the T${}_{2}$ site has two cations as NNs. The interstitial defects are illustrated in Fig. 1 in Ref.~\onlinecite{rak2014}.

\section{\label{subst_B}Substitutional B defects in NFO}

In the primary coolant of a PWR, B is present in the form of boric acid (H${}_{3}$BO${}_{3}$). Therefore, the process of B incorporation into NFO as a substitutional impurity can be envisioned as the addition of one H${}_{3}$BO${}_{3 }$ molecule to the NFO followed by the removal of a cation (Fe${}^{3+}$ or Ni${}^{2+}$) and three hydroxyl (OH${}^{-}$) ions from the NFO supercell. This process can be described by the reaction:

\begin{equation} \label{eq12}
NFO+H_{3} BO_{3} \to NFO(B)+Fe^{3+} /Ni^{2+} +3(OH^{-} )
\end{equation}

To calculate the formation energies of a substitutional defect, Eq. \eqref{eq3} can be applied with $n_{H_{3} BO_{3} } =-1$, $n_{Fe^{3+} /Ni^{2+} } =1$ and $n_{OH^{-} } =3$.

Because the method for the formation energy calculation is similar for all B-related defects, we will describe the details for the susbtitutional B at Fe site $\left(B_{Fe^{T/O} } \right)$, and for the other defects we only present the results. In the case of $B_{Fe} $, taking into consideration that the ionic charges are $z_{Fe^{3+} } =3$, $z_{OH^{-} } =-1$, and  $z_{H_{3} BO_{3} } =0$, the fourth term on the right side of Eq.~\eqref{eq3} cancels out, indicating that the charge neutrality of the solution is maintained after the substitution takes place. This makes sense because in this process one Fe${}^{3+}$ ion in NFO is replaced by one B${}^{3+}$ ion. Thus the expression for defect formation energy can be written as:

\begin{equation} \label{eq13}
\Delta H_{f} \left(B_{Fe^{T/O} }^{q} \right)=\Delta E\left(B_{Fe^{T/O} }^{q} \right)+\mu _{Fe^{3+} } +3\mu _{OH^{-} } -\mu _{H_{3} BO_{3} } +q\left(E_{F} +E_{VBM}^{def} \right)
\end{equation}

As described in Section~\ref{aqueous_chem_pot}, the chemical potential of aqueous species can be expressed as the sum of the standard chemical potential and a temperature dependent term. The standard chemical potential of aqueous Fe${}^{3+ }$can be readily evaluated using Eq. \eqref{eq7}. In a similar way, considering the reactions $H_{2}+1/2 O_{2} \to OH^{-} +H^{+} $ and $B+3/2 H_{2} +3/2 O_{2} \to H_{3} BO_{3} $, the standard chemical potentials of OH${}^{- }$and H${}_{3}$BO${}_{3}$ can be represented as:

\begin{equation} \label{eq14}
\mu _{OH^{-} }^{0} =\Delta G_{f}^{0} \left(OH^{-} \right)+\mu _{O}^{0} +\mu _{H_{2} }^{0} -\mu _{H^{+} }^{0}
\end{equation}

\begin{equation} \label{eq15}
\mu _{H_{3} BO_{3} }^{0} =\Delta G_{f}^{0} \left(H_{3} BO_{3} \right)+\mu _{B,s}^{0} +\frac{3}{2} \mu _{H_{2} }^{0} +\frac{3}{2} \mu _{O}^{0}
\end{equation}
Inserting the expressions of the standard chemical potentials (Eqs. \eqref{eq14} and \eqref{eq15}) into Eq. \eqref{eq13}, the formation energy of substitutional B at an Fe site can be written as:

\begin{equation} \label{eq16}
\begin{array}{rcl} {\Delta H_{f} \left(B_{Fe^{T/O} }^{q} \right)} & {=} & {\Delta E\left(B_{Fe^{T/O} }^{q} \right)+\Delta G_{f}^{0} \left(Fe^{3+} \right)+3\Delta G_{f}^{0} \left(OH^{-} \right)-\Delta G_{f}^{0} \left(H_{3} BO_{3} \right)+\mu _{Fe,s}^{0} -} \\ {} & {} & {-\mu _{B,s}^{0} +RT\left(\ln \left[Fe^{3+} \right]+3\ln \left[OH^{-} \right]-\ln \left[H_{3} BO_{3} \right]\right)+q\left(E_{F} +E_{VBM}^{def} \right)} \end{array}
\end{equation}
To calculate the formation energy, as expressed by Eq. \eqref{eq16}, the concentrations of H${}_{3}$BO${}_{3 }$and OH${}^{-}$are needed. The former is approximated by the experimental value of 1400 ppm B (2.26$\times$10${}^{-2}$ mol/dm${}^{3}$) at the beginning of the cycle~\cite{epri_3, epri_4} and the latter is estimated from the ionization constant of water, $pK_{w} =-\log \left(\left[H^{+} \right]\left[OH^{-} \right]\right)$. Using a semi-empirical equation~\cite{bandura2006}, \textit{pK${}_{w}$} is calculated for a temperature range of 25 to 350${}^\circ$ C and is combined with the $pH=-\log \left[H^{+} \right]$ of the primary coolant to obtain the hydroxyl concentration as $\left[OH^{-} \right]=10^{pH-pK_{w} } $. This approach allows us to include the pH dependence into the calculations. Equation \eqref{eq16} indicates that the formation energy depends on the charge state of the defect (q) and the energy of the electron reservoir (Fermi level, E${}_{F}$). Both of these quantities are kept as parameters, with q ranging from -2 to 2 and with E${}_{F }$ taking values within the band gap of NFO. In the present work we employ the electronic structure of NFO as calculated in Ref.~\cite{rak2014}; thus the allowed values of E${}_{F }$ range from zero, corresponding to energy of the valence band maximum (VBM), to 1.3 eV, corresponding to the conduction band minimum (CBM).

To calculate the energy required to incorporate B at a Ni site,  Eq. \eqref{eq3} can be applied with $n_{H_{3} BO_{3} } =-1$, $n_{Ni^{2+} } =1$, $n_{OH^{-} } =3$ and $z_{H_{3} BO_{3} } =0$, $n_{Ni^{2+} } =2$, $n_{OH^{-} } =-1$:

\begin{equation} \label{eq17}
\Delta H_{f} \left(B_{Ni}^{q} \right)=\Delta E\left(B_{Ni}^{q} \right)+\mu _{Ni^{2+} } +3\mu _{OH^{-} } -\mu _{H_{3} BO_{3} } +\left(E_{F} +E_{VBM}^{0} \right)+q\left(E_{F} +E_{VBM}^{def} \right)
\end{equation}
The main difference between formation energies of $B_{Fe} $ and$B_{Ni} $, as described by Eqs. \eqref{eq13} and \eqref{eq17}, is related to the presence of the extra term $\left(E_{F} +E_{VBM}^{0} \right)$in the latter. This term is necessary to restore the charge neutrality of the process wherein Ni${}^{2+ }$ in NFO is replaced by B${}^{3+}$.

\begin{figure}[t]
\centering
\includegraphics[scale=0.45]{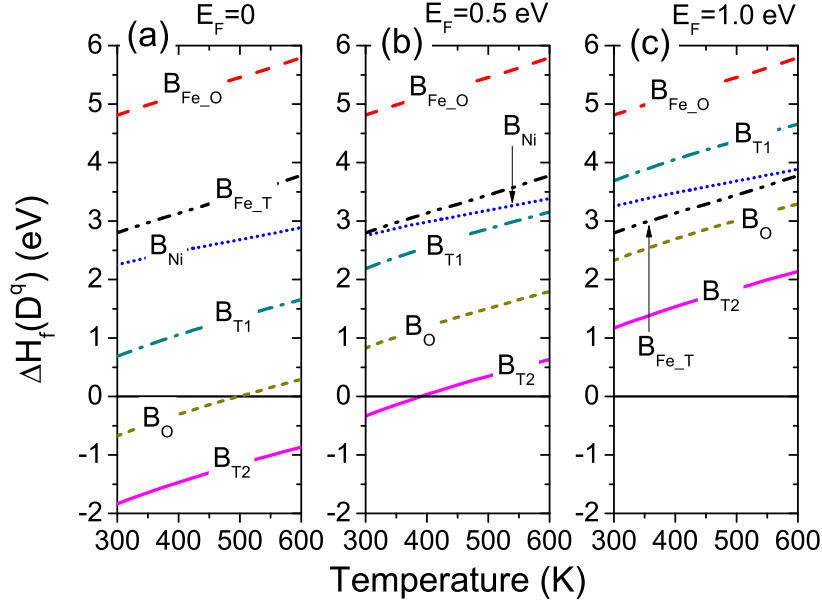}
\caption{(Color online) Temperature dependence of the B-related defect formation energies in NFO, as calculated for (a) E${}_{F}$ = 0, (b) E${}_{F}$ = 0.5 eV, and (c) E${}_{F}$ = 1.0 eV. The most stable defects are interstitial B at  tetrahedral and octahedral sites.}
\label{fig2}
\end{figure}

Applying the method described above, the formation energy of a substitutional B at a Ni site in NFO can be written as:
\begin{equation} \label{eq18)}
\begin{array}{rcl} {\Delta H_{f} \left(B_{Ni}^{q} \right)} & {=} & {\Delta E\left(B_{Ni}^{q} \right)+\Delta G_{f}^{0} \left(Ni^{2+} \right)+3\Delta G_{f}^{0} \left(OH^{-} \right)-\Delta G_{f}^{0} \left(H_{3} BO_{3} \right)+} \\ {} & {} & {+\mu _{Ni,s}^{0} -\mu _{B,s}^{0} -\left(\mu _{H^{+} }^{0} -{1\mathord{\left/ {\vphantom {1 2}} \right. \kern-\nulldelimiterspace} 2} \mu _{H_{2} }^{0} \right)+\left(E_{F} +E_{VBM}^{0} \right)+q\left(E_{F} +E_{VBM}^{def} \right)+} \\ {} & {} & {+RT\left(\ln \left[Ni^{2+} \right]+3\ln \left[OH^{-} \right]-\ln \left[H_{3} BO_{3} \right]\right)} \end{array}
\end{equation}
The formation energies of neutral B defects (q=0) as a function of temperature, calculated for pH = 7 and E${}_{F}$ = 0, 0.5, and 1.0 eV are illustrated in Fig.~\ref{fig2} (a), (b), and (c), respectively. A general trend, characteristic to all defects, is that the formation energies increase with temperature, indicating that B incorporation in NFO becomes more energetically favorable as the coolant temperature decreases during plant shutdown. As regards to the substitutional defects, in the entire temperature range, the formation energies are relatively high, suggesting that the incorporation of B in NFO as a substitutional defect is unlikely. As illustrated in Fig.~\ref{fig2} (b) and (c), the high formation energy values are maintained for all values of E${}_{F}$ within the bandgap. In the case of neutral defects (q=0), any change in the position of E${}_{F}$ only affects the formation of energy of $B_{Ni}$ while the energies of $B_{Fe^{T} }$ and $B_{Fe^{O} }$ remain unchanged. This is because, unlike $B_{Fe^{T} } $ and $B_{Fe^{O} }$, the formation of $B_{Ni}$ requires electron exchange with the electron reservoir to maintain the neutrality of the substitution process.  As evident from Eq. \eqref{eq17}, as the E${}_{F}$ value increases (Fermi level closer to CBM) more energy is needed to add electrons to the electron reservoir, therefore the formation energy of $B_{Ni} $ increases.

\section{\label{interst_B}Interstitial B defects in NFO }

\begin{figure}[t]
\centering
\includegraphics[scale=0.35]{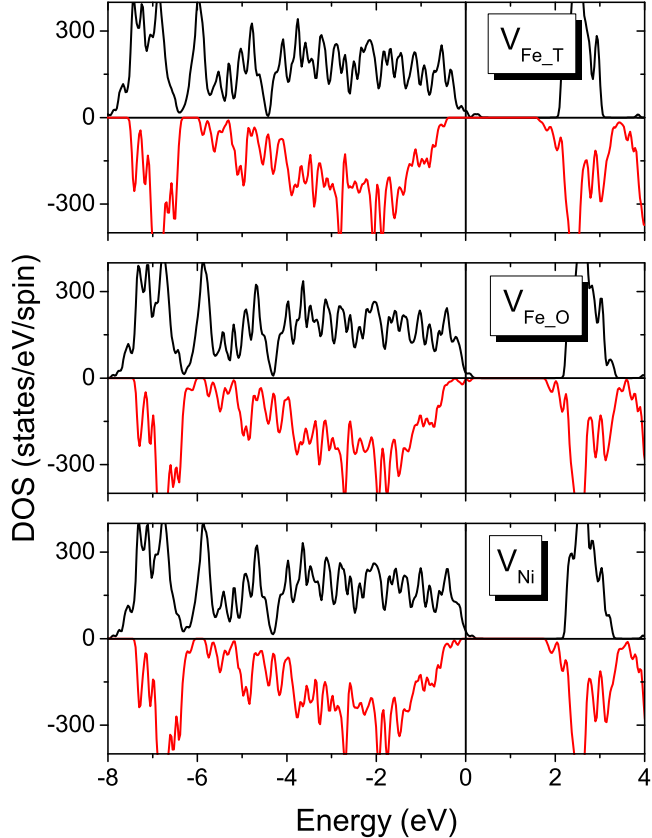}
\caption{(Color online) Total DOS calculated for vacancy containing NFO. All three defects (tetrahedral Fe vacancy -- upper panel, octahedral Fe vacancy -- middle panel, and Ni vacancy -- lower panel) introduce defect states at the top of the VBM.}
\label{fig3}
\end{figure}

\begin{figure}[t]
\centering
\includegraphics[scale=0.45]{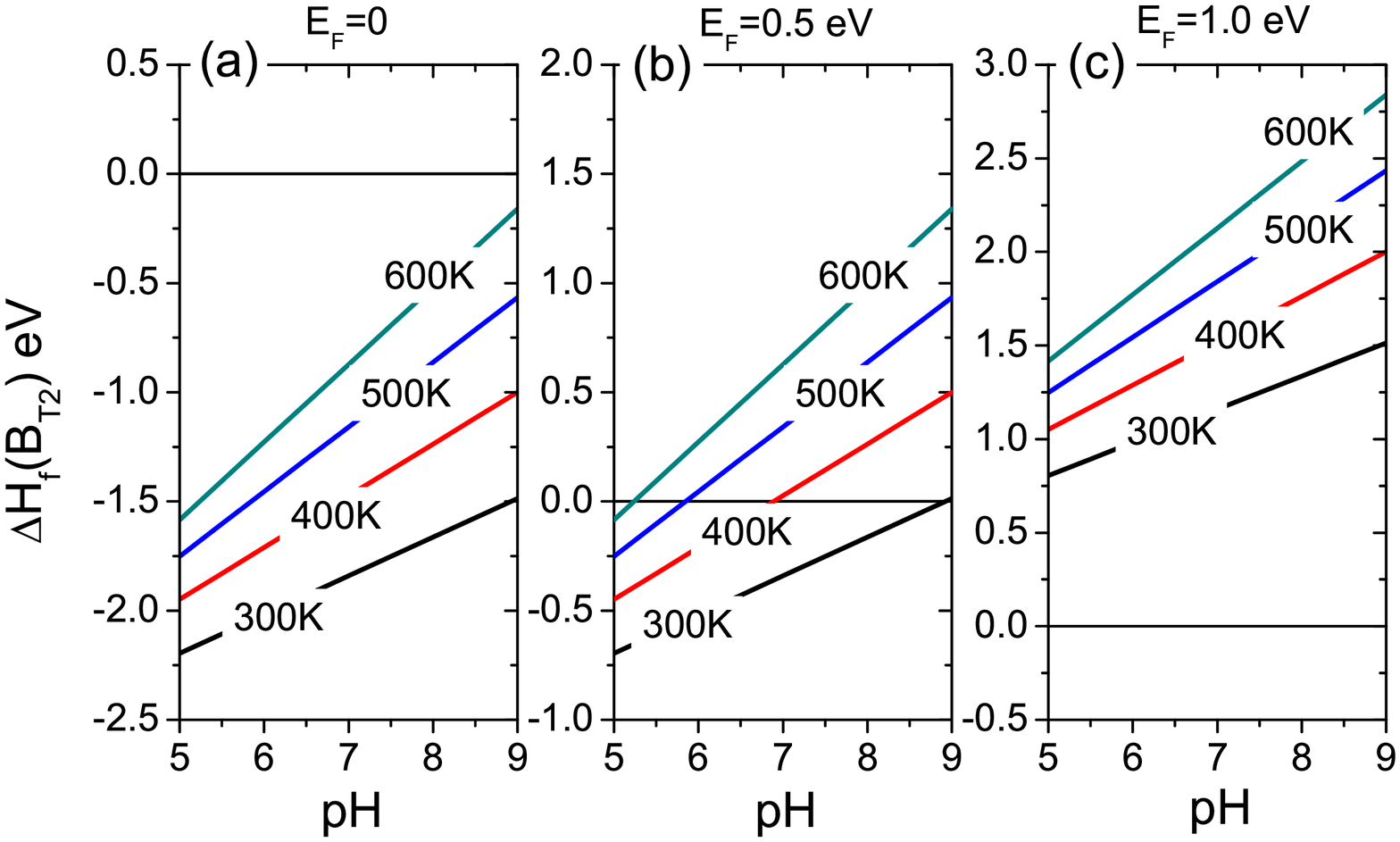}
\caption{(Color online) pH dependence of  the formation energy of$B_{T_{2} } $, calculated at various temperatures and E${}_{F }$= 0.5 eV. At higher pH values the incorporation of B into NFO becomes less energetically favorable.}
\label{fig4}
\end{figure}

The process of B incorporation into NFO as an interstitial defect can be described as the addition of one H${}_{3}$BO${}_{3}$ molecule to NFO followed by the removal of three OH${}^{-}$${}_{ }$groups:

\begin{equation} \label{eq19}
NFO+H_{3} BO_{3} \to NFO\left(B\right)+3\left(OH^{-} \right)
\end{equation}

The formation energies can be calculated by employing Eq. \eqref{eq3} with $n_{H_{3} BO_{3} } =-1$ and $n_{OH^{-} } =3$:

\begin{equation} \label{eq20}
\begin{array}{rcl} {\Delta H_{f} \left(B_{T_{1} /T_{2} /O}^{q} \right)} & {=} & {\Delta E\left(B_{T_{1} /T_{2} /O}^{q} \right)+3\Delta G_{f}^{0} \left(OH^{-} \right)-\Delta G_{f}^{0} \left(H_{3} BO_{3} \right)-\mu _{B,s}^{0} -} \\ {} & {} & {-3\left(\mu _{H^{+} }^{0} -\frac{1}{2} \mu _{H_{2} }^{0} \right)+3\left(E_{F} +E_{VBM}^{0} \right)+q\left(E_{F} +E_{VBM}^{def} \right)+} \\ {} & {} & {+RT\left(3\ln \left[OH^{-} \right]-\ln \left[H_{3} BO_{3} \right]\right)}
\end{array}
\end{equation}

The calculated values for the neutral (q=0) interstitial defects as a function of temperature, for pH = 7 and E${}_{F }$= 0, 0.5, and 1 eV, are illustrated in Fig.~\ref{fig2} (a), (b), and (c). The formation energies exhibit a strong dependence on the E${}_{F }$ value. This is also evident from Eq. \eqref{eq20}, where the term $\left(E_{F} +E_{VBM}^{0} \right)$ is multiplied by a factor of three. This is because during the process of B incorporation as an interstitial impurity in NFO, three OH${}^{-}$ groups are released (see Eq. \eqref{eq19}), and therefore to maintain charge neutrality three electrons must be added to the electron reservoir. Thus, the closer the E${}_{F}$ to the CBM, the higher the formation energy of the interstitial defects. Nevertheless, for all values of E${}_{F}$ within the NFO bandgap, the incorporation of B as an interstitial defect ($B_{T_{2} } $and $B_{O} $) is more energetically favorable than the creation of any substitutional B defect. The most remarkable feature, however, is that for values of E${}_{F}$ below the midgap $\left(E_{F} \le 0.5{\rm \; eV}\right)$ the formation energies of $B_{T_{2} }$ and $B_{O}$ become negative, implying that B can be incorporated in substantial quantities in p-type NFO. Furthermore, at lower temperatures, the formation energies decrease considerably, suggesting that during the shutdown of a PWR, the absorbed B becomes energetically more stable within the CRUD.

Experimentally it has been observed that NFO accommodates nonstoichiometry with Fe/Ni ratios above and below 2.0~\cite{paladino1959, obryan1965}. More recently, investigations of thermophysical properties of NFO in the temperature range of 300-1300K also revealed that minor nonstoichiometry is responsible for variations in the Curie temperature relative to previously reported literature values~\cite{nelson2014}. To assess the electrical properties (n-type or p-type) associated with the nonstoichiomentry, we examine the electronic structure of vacancy-containing NFO. Specifically, we look at the nature of defect states introduced by tetrahedral and octahedral Fe vacancies $\left(V_{Fe^{T} } \right. $ and $\left. V_{Fe^{O} } \right)$ and Ni vacancy $\left(V_{Ni} \right)$ in NFO. From the calculated density of states (DOS), illustrated in Fig.~\ref{fig3}, it is apparent that all three vacancies introduce acceptor states right above the VBM, contributing to p-type conductivity in NFO. Because incorporation of B is favored in p-type NFO, this finding reinforces the idea that the B added to the primary coolant might be absorbed by the CRUD as interstitial impurities.

One method to control corrosion in a PWR is to tune the pH of the primary coolant. Because the most stable defect in our study is the interstitial B $\left(B_{T_{2} } \right)$, we analyze the pH dependence of its formation energy. This is illustrated in Fig.~\ref{fig4} for various temperatures and three different values of $E_{F} $.  It is noticeable that the formation energy of $B_{T_{2} } $ increases strongly with pH, suggesting that a slightly basic coolant might prevent B uptake by CRUD as an interstitial defect. However, in the case of strongly p-type NFO, as illustrated in Fig.~\ref{fig4} (a), even at elevated pH the formation energy of $B_{T_{2} } $is negative, indicating that B is energetically stable inside the crystal structure of the CRUD.

\section{\label{summary}Summary}

An innovative approach that combines first-principles calculations with thermodynamic data, has been used to evaluate formation energies of B-related defects in NFO assuming chemical equilibrium with aqueous solutions. The approach has the advantage that the ionic charge is decoupled from the charge state of the defect. Charge neutrality is achieved through exchange of electrons with the electron reservoir that has energy equal to the Fermi level. This allows for the investigation of defects whose charge states are different from the ionic charge added to or removed from the system. Furthermore, the scheme extends the 0K DFT results to higher temperatures and pressures and includes pH and concentration dependence.

The method has been employed to investigate the energetics of B stability in NFO as a substitutional or interstitial impurity. Calculations have been carried out assuming solid-liquid equilibrium between NFO and an aqueous solution at conditions of temperature, pressure, and Ni${}^{2+}$, Fe${}^{3+}$ and H${}_{3}$BO${}_{3}$ concentrations that are characteristic to the primary coolant of a PWR. The results indicate that in the temperature range of 300 to 600 K, the interstitial B impurities are thermodynamically more stable than the substitutional defects. The formation energies exhibit a strong dependence on the position of the Fermi level (E${}_{E}$) within the bandgap of the host NFO, p-type NFO being more favorable for B incorporation. Analysis of the electronic DOS associated with vacancies in NFO indicate that both Fe and Ni vacancies generate p-type conductivity, suggesting that nonstoichiometric NFO, that is predominantly present in PWR CRUD, is favorable for accommodating B interstitials. The examination of the pH dependence of the defect stability indicates that a basic PWR coolant might be necessary to mitigate the B uptake by the CRUD. The results of the present investigation reveal that under operating PWR conditions, B is stable in NFO as an interstitial impurity, therefore it can accumulate in the atomic structure of CRUD and can be a possible large contributor to AOA.

\begin{acknowledgements}

\noindent This research was supported by the Consortium for Advanced Simulation of Light Water Reactors (CASL, http://www.casl.gov), an Energy Innovation Hub (http://www.energy.gov/hubs) for Modeling and Simulation of Nuclear Reactors under U.S. Department of Energy Contract No. DE-AC05-00OR22725. The computational work has been performed at NERSC, supported by the Office of Science of the US Department of Energy under Contract No. DE-AC02-05CH11231.

\end{acknowledgements}

\appendix*
\section{Elemental-phase chemical potentials}

\begin{table}[h]
\caption{\label{table2} Calculated chemical potentials of elemental substances in their conventional reference phase.}
\begin{ruledtabular}
\begin{tabular}{l c l c l c}
Element & $\mu _{M,solid}^{0} $ (eV) & Element & $\mu _{M,solid}^{0} $ (eV) & Element & $\mu _{M,solid}^{0} $ (eV) \\ \hline
Ag & -0.86 & Ge & -4.21 & Pt & -3.97 \\
Al & -3.12 & Hf & -7.55 & Rb & -0.70 \\
As & -4.92 & Hg & -0.18 & Rh & -4.79 \\
Au & -2.27 & In & -2.39 & S & -3.99 \\
\textbf{B} & \textbf{-6.56} & Ir & -5.99 & Sb & -4.24 \\
Ba & -1.40 & K & -0.83 & Sc & -4.71 \\
Be & -3.45 & La & -3.72 & Se & -3.44 \\
Bi & -4.27 & Li & -1.68 & Si & -5.12 \\
Ca & -1.70 & Mg & -1.11 & Sn & -3.85 \\
Cd & -0.66 & Mn & -6.85 & Sr & -1.20 \\
Cl & -1.60 & N & -8.37 & Ta & -8.96 \\
Co & -4.79 & Na & -1.09 & Te & -3.18 \\
Cr & -7.14 & Nb & -6.91 & Ti & -5.59 \\
Cu & -2.00 & \textbf{Ni} & \textbf{-3.69} & V & -6.49 \\
F & -1.68 & \textbf{O} & \textbf{-4.71} & Y & -4.90 \\
\textbf{Fe} & \textbf{-6.10} & P & -5.47 & Zn & -0.93 \\
Ga & -2.51 & Pd & -3.14 & Zr & -6.02 \\
\end{tabular}
\end{ruledtabular}
\end{table}

\end{document}